\documentclass[useAMS,usenatbib]{mn2e}
\usepackage{epsfig}

\title{Star formation in metal-poor gas clouds}
\author[Glover \& Clark]
{Simon C. O. Glover\thanks{E-mail: glover@uni-heidelberg.de} \& Paul C. Clark \\
\\ Institut f\"ur Theoretische Astrophysik, Zentrum f\"ur Astronomie  der Universit\"at Heidelberg, Albert-Ueberle-Stra\ss e 2,  \\ 69120 Heidelberg, Germany
}

\begin{document}
\maketitle

\begin{abstract}
Observations of molecular clouds in metal-poor environments typically find that they
have much higher star formation rates than one would expect based on their observed
CO luminosities and the molecular gas masses that are inferred from them. This finding
can be understood if one assumes that the conversion factor between CO luminosity 
and H$_{2}$ mass is much larger in these low metallicity systems than in nearby 
molecular clouds. However, it is unclear whether this is the only factor at work, or 
whether the star formation rate of the clouds is directly sensitive to the metallicity of
the gas.

To investigate this, we have performed numerical simulations of the coupled dynamical, 
chemical and thermal evolution of model clouds with metallicities ranging from 
$0.01 \: {\rm Z_{\odot}}$ to ${\rm Z_{\odot}}$. We find that the star formation rate in our
model clouds has little sensitivity to the metallicity. Reducing the metallicity of the gas
by two orders of magnitude delays the onset of star formation in the clouds by no more
than a cloud free-fall time and reduces the time-averaged star formation rate by at most
a factor of two. On the other hand, the chemical state of the clouds is highly sensitive to
the metallicity, and at the lowest metallicities, the clouds are completely dominated by
atomic gas. Our results confirm that the CO-to-H$_{2}$ conversion factor in these systems
depends strongly on the metallicity, but also show that the precise value is highly 
time-dependent, as the integrated CO luminosity of the most metal-poor clouds is 
dominated by emission from short-lived gravitationally collapsing regions. Finally, we find evidence 
that the star formation rate per unit H$_{2}$ mass increases with decreasing metallicity,
owing to the much smaller H$_{2}$ fractions present in our low metallicity clouds.
\end{abstract}

\begin{keywords}
galaxies: ISM -- ISM: clouds -- ISM: molecules --  stars: formation
\end{keywords}

\section{Introduction}
High resolution observations of the gas and young stars within nearby massive spiral galaxies
have shown that there is a tight correlation between the surface density of star formation and
the surface density of molecular gas within these galaxies 
\citep[see e.g.][]{wong02,leroy08,bigiel08,bigiel11,schruba11}.
However, similar observations of the star formation rate and the molecular gas surface
density in low-metallicity environments, such as dwarf galaxies, tell a rather different story. 
Carbon monoxide (CO), the most commonly-used tracer of molecular gas, is difficult to detect 
in low metallicity gas, and even when detected, the ratio of the observed star formation rate to 
the CO luminosity is much larger than in higher metallicity regions 
\citep[see e.g.][]{taylor98,leroy07,schruba11,schruba12}. If one assumes that the conversion factor 
between CO luminosity and H$_{2}$ mass
-- the so-called `X-factor' -- is the same in these systems as in the Milky Way, then these 
observations imply that there is a much larger amount of star formation per unit of molecular
gas in low metallicity regions than in the Milky Way.  Observations of larger ``main-sequence''
star-forming galaxies at $z > 1$ by \citet{genzel12} also show that the ratio of the star-formation 
rate to the CO luminosity increases with decreasing metallicity, leading to a similar conclusion.

These results are the opposite of what one would naively expect, given the crucial role that 
metals and dust play in regulating the temperature of star-forming clouds in the local ISM.
If, as is widely believed, the presence of large amounts of cold gas is a 
prerequisite for star formation, then one would expect star formation to become
less effective as the metallicity is reduced, owing to the loss of coolants from the gas, as well
as the reduced effectiveness of dust shielding. Indeed, recent work by \citet{kd11} has shown
that galaxy formation models in which the star formation efficiency decreases with decreasing
metallicity do a much better job of describing the specific star formation rates of high redshift
galaxies than models in which the star formation rate is independent of metallicity.

One way to avoid this apparent contradiction is to adopt an X-factor with a strong dependence
on metallicity. In this picture, one assumes that the CO luminosity per unit H$_{2}$ mass in
low-metallicity environments is much smaller than in the Milky Way, implying that the
actual reservoir of molecular gas present in such regions is much larger than one would
predict from applying the Galactic X-factor to the observed CO emission. In other words, rather 
than having anomalously large star formation efficiencies, low metallicity dwarf galaxies
and other similar systems may simply
be strongly deficient in CO. This idea has considerable theoretical support from 
numerical models of the dependence of the X-factor on metallicity \citep[see e.g.][]{bell06,gm11,
narayanan11,feld12}, and also some observational support from direct measurements of the 
X-factor in the low metallicity environment of the Small Magellanic Cloud \citep{leroy08,bol11}.

If this model is correct, then it implies that the observations are not giving us an accurate 
picture of the relationship between the gas surface density and the star formation rate in
these low-metallicity systems, and that the question of the star formation efficiency of the
gas remains unresolved. Therefore, in order to explore the extent to which the star formation 
rate within a typical gravitationally-bound molecular cloud depends on the metallicity of the
cloud, we have performed a series of numerical simulations of the chemical, thermal and 
dynamical evolution of model clouds with a wide range of metallicities. The layout of the
remainder of our paper is as follows. In Section 2, we describe the approach used for 
our numerical simulations and the initial conditions that we adopt. In Section 3, we present
our main results, and in Section 4, we discuss what these results imply regarding the 
dependence of the star formation rate on metallicity. Finally, we give our conclusions in
Section 5.

\section{Simulations}
\label{sims}
\subsection{Numerical method}
The simulations described in this paper were performed using a modified version of the Gadget 2 
smoothed particle hydrodynamics (SPH) code \citep{springel05}. Our modifications include a sink 
particle algorithm for treating gravitationally collapsing regions that become too small to resolve 
\citep{bbp95}, the inclusion into the equation of motion of an optional confining pressure term 
\citep{benz90}, a treatment of gas-phase chemistry (described in more detail below), as well as 
radiative heating and cooling from a number of atomic and molecular species \citep{gj07,g10,gc12a}, and an approximate treatment of the attenuation of the Galactic interstellar radiation field (ISRF). The effects of magnetic fields are not included.

Our treatment of the gas-phase chemistry combines the hydrogen chemistry network introduced in
\citet{gm07a,gm07b} with the treatment of CO formation and destruction proposed by \citet{nl99}.
This combined network was tested against other simplified chemical networks for CO formation
and destruction by \citet{gc12a}, who showed that it does a very good job of reproducing the
results of more complex networks (e.g.\ \citealt{g10}) for C$^{+}$, C and CO, while incurring only
one-third of the computational cost. Our treatment of H$_{2}$ formation on dust grains follows
that of \citet{hm79}, {but we have verified that in the conditions studied in this paper, we would get very
similar results if we were to use a more modern treatment such as that in \citet{ct04} and 
\citet{cs04}.} Further details regarding our treatment of the chemistry can be found in \citet{gc12a}.
Note that the simulations presented in this paper do not include the effects of the freeze-out of CO
onto dust grains. In Galactic star-forming clouds, CO freeze-out has been shown to have only a 
very minor effect on the thermal balance of the gas \citep{gold01}, and we expect it to be even
less important in lower metallicity clouds, as the dust temperatures in these clouds will generally 
be somewhat higher than in the typical Galactic case.

We assume {in the majority of our simulations} that our clouds are illuminated by the standard 
interstellar radiation field (ISRF), as parameterized by \citet{dr78} in the ultraviolet and by \citet{bl94} at 
longer wavelengths. {The effects of varying the strength of the ISRF are examined in 
Section~\ref{ISRF}}. To treat dust extinction, H$_{2}$ self-shielding, CO self-shielding and the shielding of CO by 
H$_{2}$, we use the {\sc treecol} algorithm \citep{cgk11}, as described in 
\citet{gc12b}. This algorithm provides us with an approximate 4$\pi$ steradian map of the column 
densities of hydrogen nuclei, H$_{2}$ and CO seen by each SPH particle, discretized into 48 
equal-area pixels. To convert from the hydrogen column density to the dust extinction, we use
the relationship $A_{\rm V} = 5.348 \times 10^{-22} (N_{\rm H, tot} / 1 \: {\rm cm^{-2}})
({\rm Z} / {\rm Z_{\odot}})$ \citep{bsd78,db96}, where we have assumed that the extinction
scales linearly with the total metallicity. To convert from the H$_{2}$ and CO column densities
to the shielding factors due to H$_{2}$ self-shielding, CO self-shielding and the shielding of
CO by H$_{2}$, we use conversion factors from \citet{db96} for H$_{2}$ self-shielding and
\citet{lee96} for the other two terms.

\subsection{Initial conditions}
The initial state of our cloud is a uniform sphere with a total mass of 10$^4 \rm{M}_\odot$ and an
initial hydrogen nuclei number density $n = 300 \: {\rm cm}^{-3}$, giving it an initial radius of
approximately 6~pc. This cloud mass is typical of a small, nearby molecular cloud, such as
e.g.\ the Perseus or Taurus molecular clouds.
We have chosen to focus on the behaviour of a relatively small molecular cloud for two main
reasons. First, it is simpler to simulate a small cloud with the required mass resolution than it
would be to simulate a much larger cloud, since the required number of SPH particles scales
linearly with the cloud mass. Second, a small molecular cloud will typically have a larger ratio
of thermal to gravitational energy than a larger cloud, and so it is reasonable to expect that 
varying the metallicity will have more influence on the star formation rate in a small cloud than
in a large one. We show in Section~\ref{rate} below that large variations in the metallicity have only
a small effect on the star formation rate, and because we are focussing on the case where we
expect the metallicity variations to have the greatest effect, we can reasonably conclude that
they will have even less effect in larger molecular clouds.

To model the gas making up the cloud, we use two million SPH particles, giving us a particle mass
of $0.005 \: {\rm M_{\odot}}$ and a mass resolution of $0.5 \: {\rm M_{\odot}}$. We showed in 
\citet{gc12b} that this mass resolution is sufficient to accurately determine the star formation rate,
but that it does not allow us to draw strong conclusions regarding the form of the low-mass end
of the stellar initial mass function. We inject bulk (non-thermal) motions into the cloud by imposing a turbulent velocity field that has a initial power spectrum $P(k) \propto k^{-4}$, where $k$ is the
wavenumber. The energy in the turbulence is initially equal to the gravitational energy in the cloud, 
meaning that the initial root mean squared turbulent velocity is around 3 km\,s$^{-1}$. During the
simulation,  the turbulence is allowed to freely decay via shocks and compression-triggered cooling. 
We set the initial gas temperature to 20~K and the initial dust temperature to 10~K, but these values both alter rapidly once the simulation begins, as the gas and dust relax towards thermal equilibrium.

We consider five different metallicities in this study: ${\rm Z} = 0.01, 0.03, 0.1, 0.3,$ and $1.0 \: {\rm Z_{\odot}}$. 
In our solar metallicity runs, we adopt the values $x_{\rm C} = 1.4 \times 10^{-4}$ and
$x_{\rm O} = 3.2 \times 10^{-4}$ for our total oxygen and carbon abundances relative to hydrogen
\citep{sem00}. In the lower metallicity runs, we assume that $x_{\rm C}$ and $x_{\rm O}$ scale
linearly with the metallicity. We also assume that our dust abundance scales linearly with ${\rm Z}$,
but for simplicity assume that its properties remain the same (i.e.\ we do not change the form of
the extinction curve as we move to lower ${\rm Z}$, merely the overall normalization). In every
simulation, we assume that all of the available carbon starts in the form of C$^{+}$ and that all
of the available oxygen starts in neutral atomic form. For each metallicity, we perform two simulations,
one in which the hydrogen is initially fully molecular, and a second in which it is initially fully atomic.
We denote the simulations that start with molecular hydrogen by labels of the form Z$n$-M, where 
$n$  refers to the metallicity (e.g.\ Z1-M corresponds to solar metallicity, Z01-M to $0.1 \: {\rm Z_{\odot}}$, etc.), while for the simulations starting with atomic hydrogen, we use labels of the form Z$n$-A.

Finally, we note that the cosmic ray ionization rate of atomic hydrogen in {most} of our simulated clouds 
was $\zeta_{\rm H} = 10^{-17} \: {\rm s^{-1}}$. The cosmic ray ionization rates for the other major 
chemical species tracked in our chemical model were assumed to have the same ratio relative to
the rate for atomic hydrogen as given in the UMIST99 chemical database \citep{teu00}. {We 
explore the effects of varying $\zeta_{\rm H}$ in Section~\ref{cosmic}.}

\section{Results}
\subsection{Star formation rate}
\label{rate}
We first examine how the star formation rate in our model clouds varies as we vary the metallicity. 
As in \citet{gc12b}, we use 
the total mass of gas incorporated into sink particles as a proxy for the mass of stars formed.
Although the limited mass resolution of our simulations means that the total number of sinks
formed is probably not well-resolved, we have shown in previous work that our adopted
resolution is sufficient to accurately determine the star formation rate \citep{gc12b}. 

\begin{figure}
\includegraphics[width=3.2in]{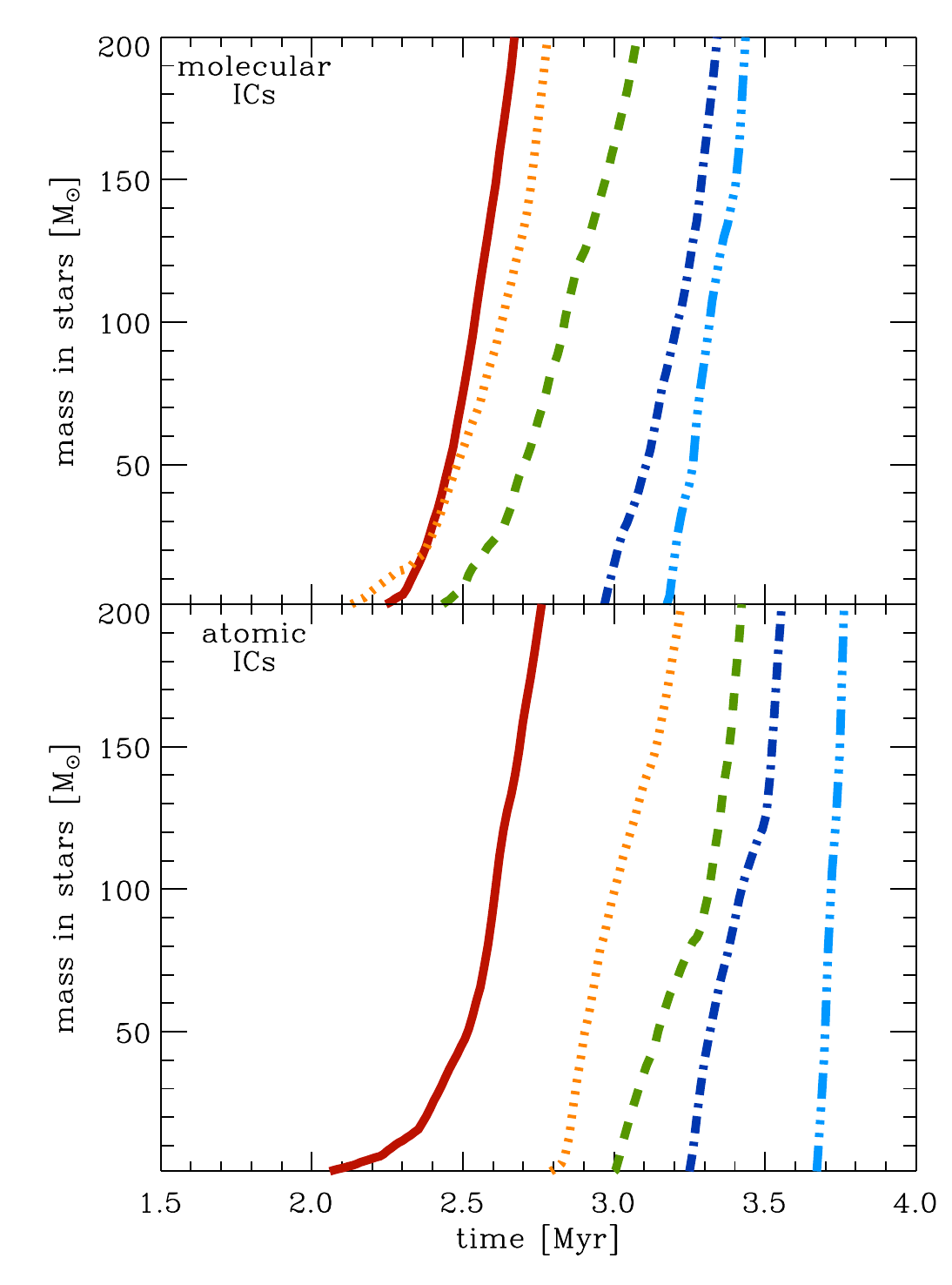}
\caption{{\it Upper panel}: Mass in sinks, plotted as a function of time, for runs Z1-M (solid line), Z03-M (dotted line), 
Z01-M (dashed line),  Z003-M (dot-dashed line) and Z001-M (dot-dot-dot-dashed line). In these runs, the hydrogen 
was initially in fully molecular form.
{\it Lower panel}: The same quantity, but for runs Z1-A (solid line), Z03-A (dotted line), Z01-A (dashed line),  
Z003-A (dot-dashed line) and Z001-A (dot-dot-dot-dashed line). In these runs, the hydrogen was initially fully 
atomic. \label{fig:sfrate}}
\end{figure}

In the upper panel of Figure~\ref{fig:sfrate}, we show how the mass in sinks varies with time in the five runs in 
which the hydrogen starts in the form of H$_{2}$, i.e.\ runs Z1-M, Z03-M, Z01-M, Z003-M and Z001-M.
In run Z1-M, the solar metallicity run, the first sink forms at $t_{\rm SF} = 2.21 \: {\rm Myr}$. For 
comparison,  the free-fall time of the cloud at its initial mean density is slightly longer, 
$t_{\rm ff} \sim 2.5 \: {\rm Myr}$. However, the turbulence in the cloud creates overdense regions 
that are able to collapse more rapidly than the cloud as a whole, and so it is not particularly surprising that
$t_{\rm SF} < t_{\rm ff}$. Once star formation has begun, it proceeds steadily, with the total
mass of stars reaching $200 \: {\rm M_{\odot}}$ after another 0.5~Myr (Table~\ref{timing}).
Measured from the start of the simulation, the star formation efficiency per free-fall time of the cloud at
this point is approximately 0.02, consistent with recent estimates of this quantity by \citet{kt07} and 
\citet{lla10}. We halt the simulation
shortly after this point because we do not include the effects of stellar feedback in our models
and hence expect them to become increasingly inaccurate as the mass incorporated into
stars increases, and as the individual stars approach the main sequence.

\begin{table}
\caption{Star formation timescales in the different runs \label{timing}}
\begin{tabular}{lcc}
\hline
Run & $t_{\rm SF}$ (Myr) & $t_{200}$ (Myr) \\
\hline
Z1-M & 2.21 & 2.67 \\  
Z1-A & 2.00 & 2.76 \\  
\hline
Z03-M & 2.11 & 2.79 \\ 
Z03-A & 2.78 & 3.22 \\ 
\hline
Z01-M & 2.42 & 3.08 \\ 
Z01-A & 2.99 & 3.41 \\ 
\hline
Z003-M & 2.95 & 3.35 \\ 
Z003-A & 3.25 & 3.56 \\ 
\hline
Z001-M & 3.17 & 3.44 \\  
Z001-A & 3.67 & 3.77 \\  
\hline
\end{tabular}
\end{table}

If we now look at what happens as we decrease the metallicity, we see that as we reduce Z, 
we delay the onset of star formation. The time at which the first sink particle forms, $t_{\rm SF}$,
increases from 2.21~Myr in the solar metallicity case to 3.17~Myr in the $0.01 \: {\rm Z_{\odot}}$
case. However, even in this case, the delay in the onset of star formation corresponds to 
considerably less than a cloud free-fall time, and star formation begins in all cases within 1.5
free-fall times. Moreover, once star formation begins, it proceeds at roughly the same rate in all 
five simulations.

The behaviour of the runs that start with fully atomic hydrogen is broadly similar. The onset of star
formation in several of these simulations is delayed compared to the fully molecular case, but 
never by more than 0.6~Myr. The spread of values for $t_{\rm SF}$ {is somewhat larger than in the
molecular case, but remains less than a factor of two}, and all of the model clouds have started forming stars 
within $1.5 \: t_{\rm ff}$ after the beginning of the simulation. It is also interesting to note that once star formation 
has begun, it tends to proceed more rapidly in the low metallicity, initially atomic clouds than in the corresponding 
runs that start with molecular gas. 

Taken together, these results demonstrate that even very large decreases in the metallicity
of the gas have only a small effect on the ability of the cloud to form stars. Star formation is
delayed for a short period in low metallicity gas compared to the solar metallicity case, but
once star formation begins, it proceeds at roughly the same rate in all of the models, to within
a  factor of a few.

\subsection{Chemical evolution}
\label{chem}
Although large changes in the metallicity of the gas have only a small effect on the 
ability of the cloud to form stars, they have a much larger effect on the chemistry of
the gas. To quantify this, it is useful to look at how the mean abundances of various
chemical species vary with time in the simulations. In our SPH simulations, the
simplest mean abundance to examine is the mass-weighted mean, defined for
a species $S$ as
\begin{equation}
\langle S \rangle_{\rm M} = \frac{1}{M} \sum_{i} m_{\rm i} x_{\rm S, i},
\end{equation}
where $m_{i}$ is the mass of particle $i$, $x_{\rm S, i}$ is the fractional abundance
of $S$ for particle $i$, $M$ is the total mass in the simulation and where we sum
over all SPH particles that have not yet been accreted by sinks. In the case of H$_{2}$, 
using this definition would yield a value of 0.5 for fully molecular gas. This is not
particularly intuitive and has the potential to cause confusion. Therefore, in this 
case alone, we use a slightly modified definition of the mean abundance which 
gives a value of 1.0 for fully molecular gas:
\begin{equation}
\langle {\rm H_{2}} \rangle_{\rm M} = \frac{1}{M} \sum_{i} 2 m_{\rm i} x_{\rm H_{2}, i},
\end{equation}

\begin{table}
\caption{Chemical state of the gas at the onset of star formation \label{chemtab}}
\begin{tabular}{lcccc}
\hline
Run & $F_{\rm H_{2}}$ & $F_{\rm C^{+}}$ & $F_{\rm C}$ & $F_{\rm CO}$ \\
\hline
Z1-M & 0.883 & 0.601 & 0.084 & 0.315 \\ 
Z1-A & 0.599 & 0.664 & 0.146 & 0.190 \\ 
\hline
Z03-M & 0.779 & 0.906 & 0.030 & 0.064 \\ 
Z03-A & 0.304 & 0.896 & 0.036 & 0.068 \\ 
\hline
Z01-M & 0.598 & 0.978 & 0.009 & 0.013 \\  
Z01-A & 0.0887 & 0.978 & 0.011 & 0.011 \\ 
\hline
Z003-M & 0.371 & 0.988 & 0.004 & 0.008 \\ 
Z003-A & 0.0191 & 0.990 & 0.004 & 0.006 \\  
\hline
Z001-M & 0.2603 & 0.990 & 0.003 & 0.007 \\ 
Z001-A & 0.00775 & 0.990 & 0.003 & 0.007 \\ 
\hline
\end{tabular}
\medskip
\\
{\bf Note:} $F_{\rm H_{2}}$ is the fraction of the total amount of hydrogen
in the form of H$_{2}$, while $F_{\rm C^{+}}$, $F_{\rm C}$ and $F_{\rm CO}$ are the 
fractions of the total available amount of carbon in the form of C$^{+}$, C and CO,
respectively.
\end{table}

\subsubsection{Carbon monoxide (CO)}
\label{co_evol}
\begin{figure}
\includegraphics[width=3.2in]{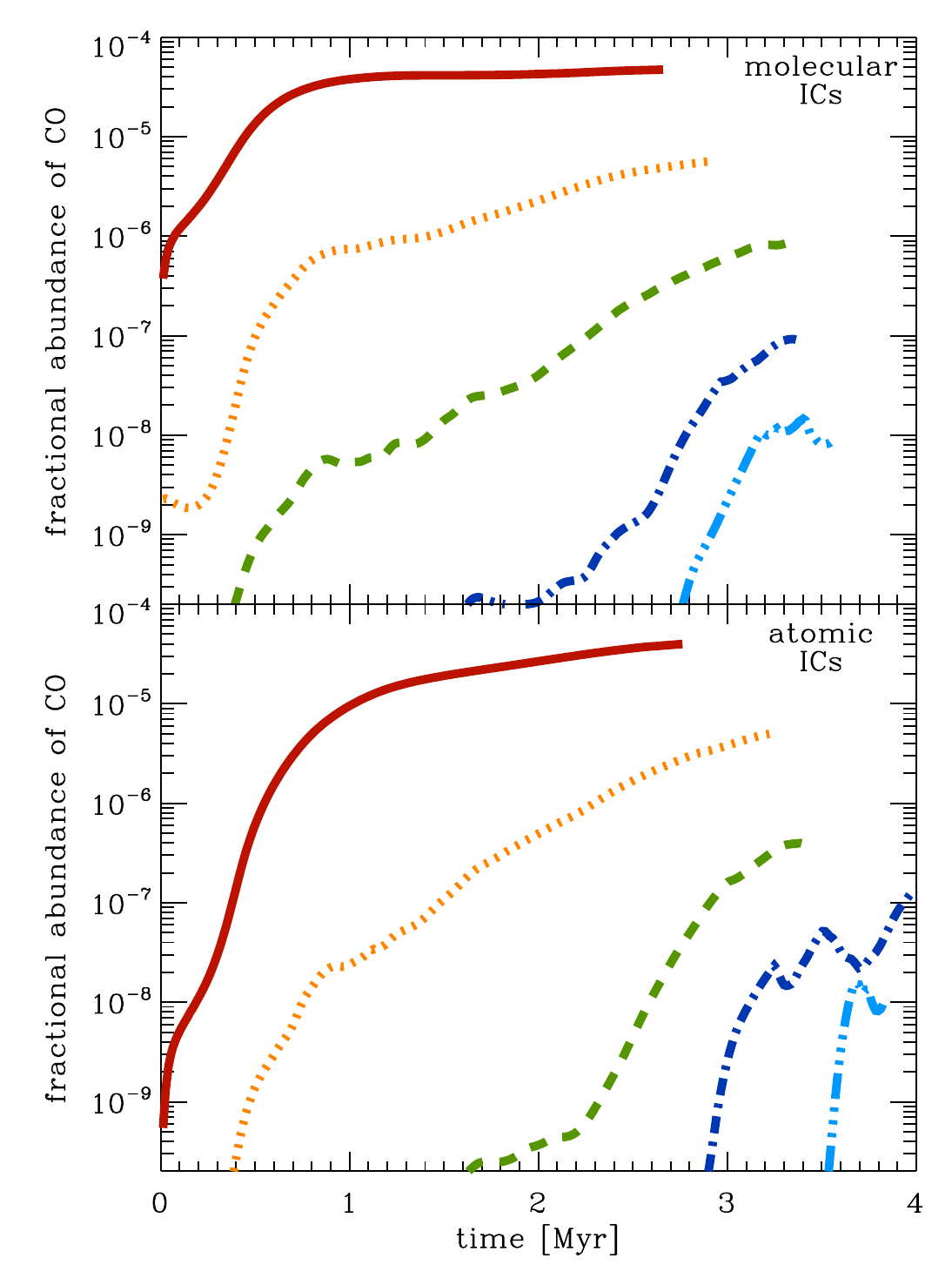}
\caption{{\it Upper panel}: Evolution with time of the mass-weighted mean fractional abundance 
of CO for runs Z1-M, Z03-M, Z01-M, Z003-M and Z001-M.
{\it Lower panel}: the same, but for runs Z1-A, Z03-A, Z01-A, Z003-A and Z001-A.
\label{fig:co-evol}}
\end{figure}

In Figure~\ref{fig:co-evol}, we show how the mass-weighted mean abundance of CO
varies with time in our simulations. At solar metallicity, our simulated clouds can
form CO relatively easily. CO formation in the two solar metallicity runs occurs 
rapidly at the beginning of the simulation, but slows down significantly by
$t \sim 1 \: {\rm Myr}$, owing to the fact that by this point, most of the dense, 
well-shielded gas that can support a high CO fraction is already fully molecular,
while the regions that remain primarily atomic have low equilibrium CO fractions.
Nevertheless, the CO fraction does not settle into a steady state, owing to the global
collapse of the cloud. This steadily increases the amount of dense, well-shielded gas
that is available, and hence allows the CO fraction to continue to increase with time,
albeit at a much slower rate.

Comparing runs Z1-M and Z1-A, we see that the initial state of the hydrogen has a
pronounced influence on the CO fraction at early times. In run Z1-M, molecular 
hydrogen is immediately available to participate in the network of chemical
reactions responsible for forming CO, allowing the CO fraction in the densest
gas to increase very quickly once the simulation begins. In run Z1-A, on the
other hand, CO formation can begin only once some H$_{2}$ has formed, and
becomes efficient only once the H$_{2}$ fraction becomes large. The growth
of the CO fraction in run Z1-A therefore lags behind that of the CO fraction in run 
Z1-M by about 0.5~Myr at early times.

As we reduce the metallicity of the clouds, it becomes much harder for them
to form large amounts of CO. The mean CO abundance decreases far more
rapidly with decreasing metallicity than would be expected simply from the
reduction in the amount of carbon available. This is illustrated particularly
clearly in Table~\ref{chemtab}, where we list the fraction of the total amount
of carbon found in the form of C$^{+}$, C and CO at the onset of star formation
in the different runs. In the solar metallicity runs, between 20--30\% of the carbon 
has been converted to CO at the point at which star formation begins. In the
$0.3 \: {\rm Z_{\odot}}$ runs, however, only 6--7\% of the available carbon is
found in the form of CO, while in the lower metallicity runs, the fraction is never
larger than around 1.5\%. 

Another important point to note from Figure~\ref{fig:co-evol} is that in the lower metallicity 
runs, most of the CO that forms does so shortly before the onset of star 
formation. As we shall see later, the CO in these runs is concentrated in 
the same high-density gas that is responsible for forming the stars
(see Section~\ref{chem_comp} below). Initially, very little of this gas exists, and the
CO fraction is very small. Once a small sub-region of the cloud becomes gravitationally
unstable and collapses, the amount of dense gas rapidly increases, as does the amount
of CO present in the cloud (although the majority of the carbon remains in the form of
C$^{+}$; see Table~\ref{chemtab}).

\subsubsection{Molecular hydrogen (H$_{2}$)}
\begin{figure}
\includegraphics[width=3.2in]{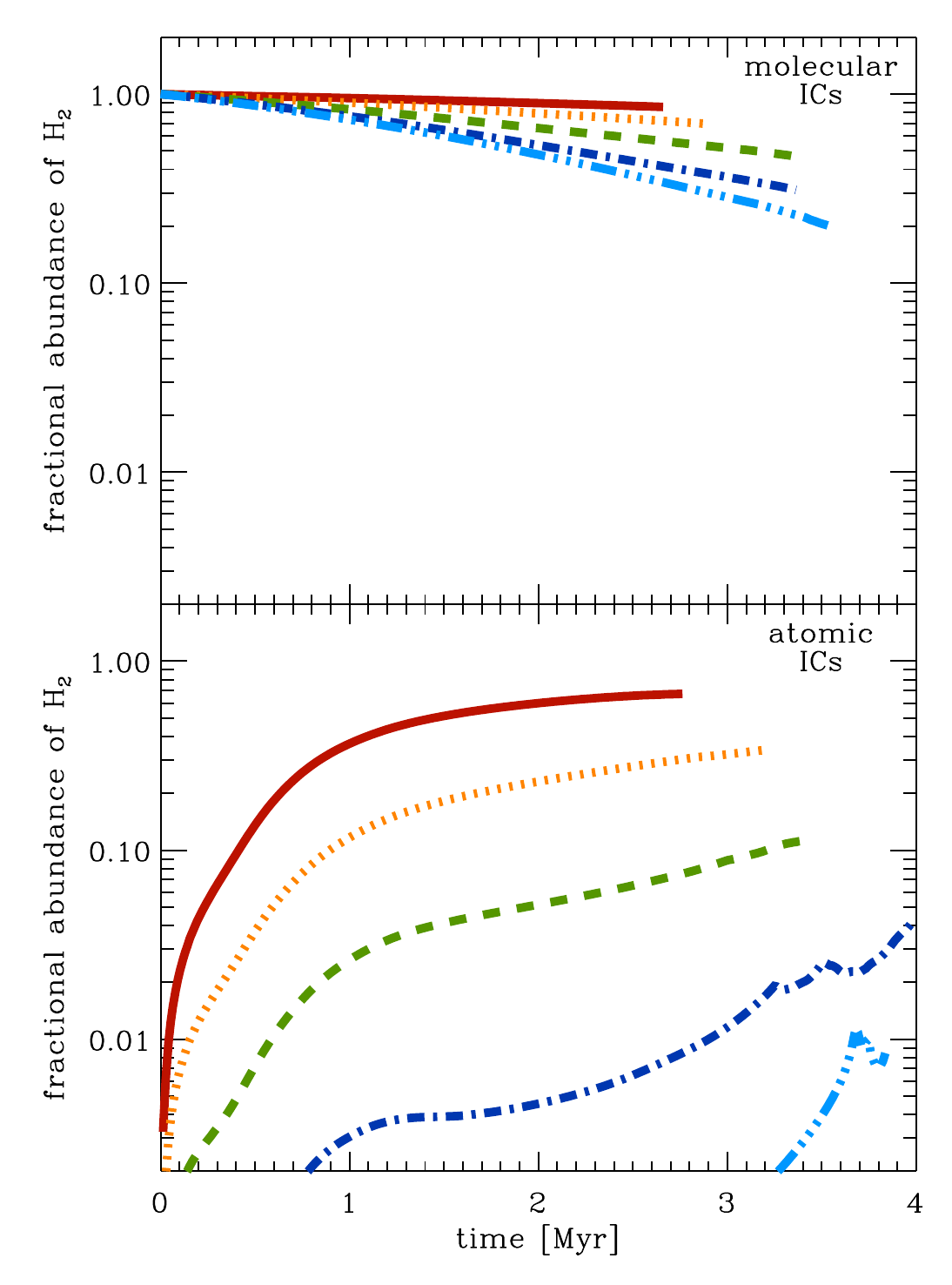}
\caption{{\it Upper panel}: Evolution with time of the mass-weighted mean fractional abundance of H$_{2}$,
for runs Z1-M, Z03-M, Z01-M, Z003-M and Z001-M.
{\it Lower panel}: the same, but for runs Z1-A, Z03-A, Z01-A, Z003-A and Z001-A. 
\label{fig:H2-evol}}
\end{figure}

Finally, it is interesting to examine how the H$_{2}$ fraction varies with time in our simulations.
This is illustrated in Figure~\ref{fig:H2-evol}. In the runs that start with the hydrogen already in the
form of H$_{2}$, we initially have more molecular gas than we would do if the cloud were in
photodissociation equilibrium, and $\langle {\rm H_{2}} \rangle_{\rm M}$ therefore decreases with 
time. Although the photodissociation timescale for unshielded H$_{2}$ is short (less than 1000~yr), 
the high level of self-shielding provided by the H$_{2}$ and the additional attenuation of the 
incoming radiation provided by the dust act to substantially reduce the H$_{2}$ photodissociation 
rate within the cloud compared to its value in unshielded gas. This has the effect of dramatically
lengthening the effective photodissociation timescale for most of the gas, with the result that
the clouds do not have sufficient time to reach chemical equilibrium before they start forming stars. 
The amount of H$_{2}$ lost from the clouds depends on the metallicity. In the solar metallicity case, 
the cloud only loses a little of its H$_{2}$, while in the runs with $Z \leq 0.03 \: {\rm Z_{\odot}}$, more
than half of the initial molecular content is destroyed, and the clouds are dominated by atomic 
hydrogen by the time that they start forming stars. 

This sensitivity to metallicity is driven by a combination of effects. As we reduce the amount of dust
in the gas, the mean extinction of the cloud decreases and so there is less attenuation of the incoming
ultraviolet radiation. At the same time, the formation rate of H$_{2}$ also decreases, as this is
directly dependent on the amount of dust present in the gas. In practice, it is the latter effect that 
appears to be more important, as demonstrated by the results from the runs that start with their
hydrogen in atomic form. In these runs, we see a strong dependence of the H$_{2}$ fraction on 
metallicity. The solar metallicity run increases its mean H$_{2}$ fraction from zero to roughly 0.6 
by the time it starts forming stars, in line with estimates from previous work \citep[see e.g.][]{gm07b,gc12b}. 
However, as we reduce the metallicity, the increase in the H$_{2}$ formation timescale makes it
much harder to produce large quantities of molecular gas even in regions that are well-shielded
from the ultraviolet radiation field. As a result, the cloud remains dominated by atomic hydrogen
in all of our runs with $Z = 0.1 \: {\rm Z_{\odot}}$ and below.

\subsection{CO distribution at the onset of star formation}
\label{chem_comp}
In Figure~\ref{fig:nt}, we show how the CO abundance of the gas varies with density and temperature
within our simulated clouds at a point just before the beginning of star formation. In every case, we 
see that at low densities, very little CO is present, while at high densities, most of the available carbon
is in the form of CO. The density at which the transition between these two regimes occurs is a strong
function of metallicity. In the solar metallicity case, it occurs at densities of around $n \sim 10^{3} \:
{\rm cm^{-3}}$, although some gas with high CO fraction can be found at lower densities, and some
higher density gas is seen to have a relatively low CO fraction. This is consistent with the idea that 
both the density and visual extinction of the gas must be high in order for it to have a high CO fraction,
and with our previous finding that there is only a poor correlation between the local volume density and 
the visual extinction of gas within turbulent clouds \citep{g10}. The clear correlation between the gas 
temperature and the CO abundance is also consistent with the extinction playing an important role: 
the gas with $T > 20 \: {\rm K}$ has a low visual extinction and consequently is heated efficiently by
photoelectric emission from dust grains, while regions with $T < 20 \: {\rm K}$ typically have higher
visual extinctions and hence are less affected by photoelectric heating.

As we move to lower metallicity, the density separating the low CO and high CO regimes 
systematically increases, evolving approximately as $n \propto {\rm Z^{-1}}$. This behaviour
is again due to the dependence of the CO abundance on the extinction of the gas. As the
metallicity decreases, a higher column density of gas is required in order to produce the
same visual extinction, and the requisite column densities tend to only be found within denser
regions of the flow. One important consequence of this behaviour is that the nature of the
CO-rich regions changes as we move to lower metallicities. In the solar metallicity case, a density
of $n \sim 10^{3} \: {\rm cm^{-3}}$ corresponds to an overdensity of only a factor of a few relative to
the initial mean density of the cloud. Such an overdensity is easily produced simply by turbulent
compression of the gas, without the need to invoke gravity, and much of the CO is located in 
overdense regions that are not self-gravitating. At lower metallicities, however, the required density
becomes much higher and therefore much harder to reach simply through turbulent compression
of the gas. Consequently, gravity begins to play more of a role, and the CO in the cloud becomes
increasingly concentrated within strongly self-gravitating, collapsing regions. This naturally explains
the strong time dependence in the CO abundance that we have already noted in Section~\ref{co_evol}.

The importance of gravitational collapse for producing the CO abundances that we find in our low
metallicity simulations can also be clearly seen if we compare our results to those from \citet{gm11}. 
This earlier study considered CO formation in turbulent gas {\em without} self-gravity, and found
that the mean CO abundance was highly sensitive to the mean visual extinction of the cloud, falling 
by almost five orders of magnitude as the mean extinction decreased from $\bar{A}_{\rm V} \sim 3$
to $\bar{A}_{\rm V} \sim 0.3$. If we consider
our model clouds at an early time, say $t \sim 1 \: {\rm Myr}$, when we expect the turbulent velocity
field to have created significant amounts of dense sub-structure but when run-away gravitational
collapse has not yet begun, then we find a similarly strong 
sensitivity to metallicity, particularly in the runs starting with atomic rather than molecular hydrogen.
If we consider much later times, however, once gravitational collapse has begun in all of the models,
then we see that the mean CO abundance becomes considerably less sensitive to the metallicity
of the cloud. 

\subsection{Temperature distribution}
\begin{figure}
\includegraphics[width=3.2in]{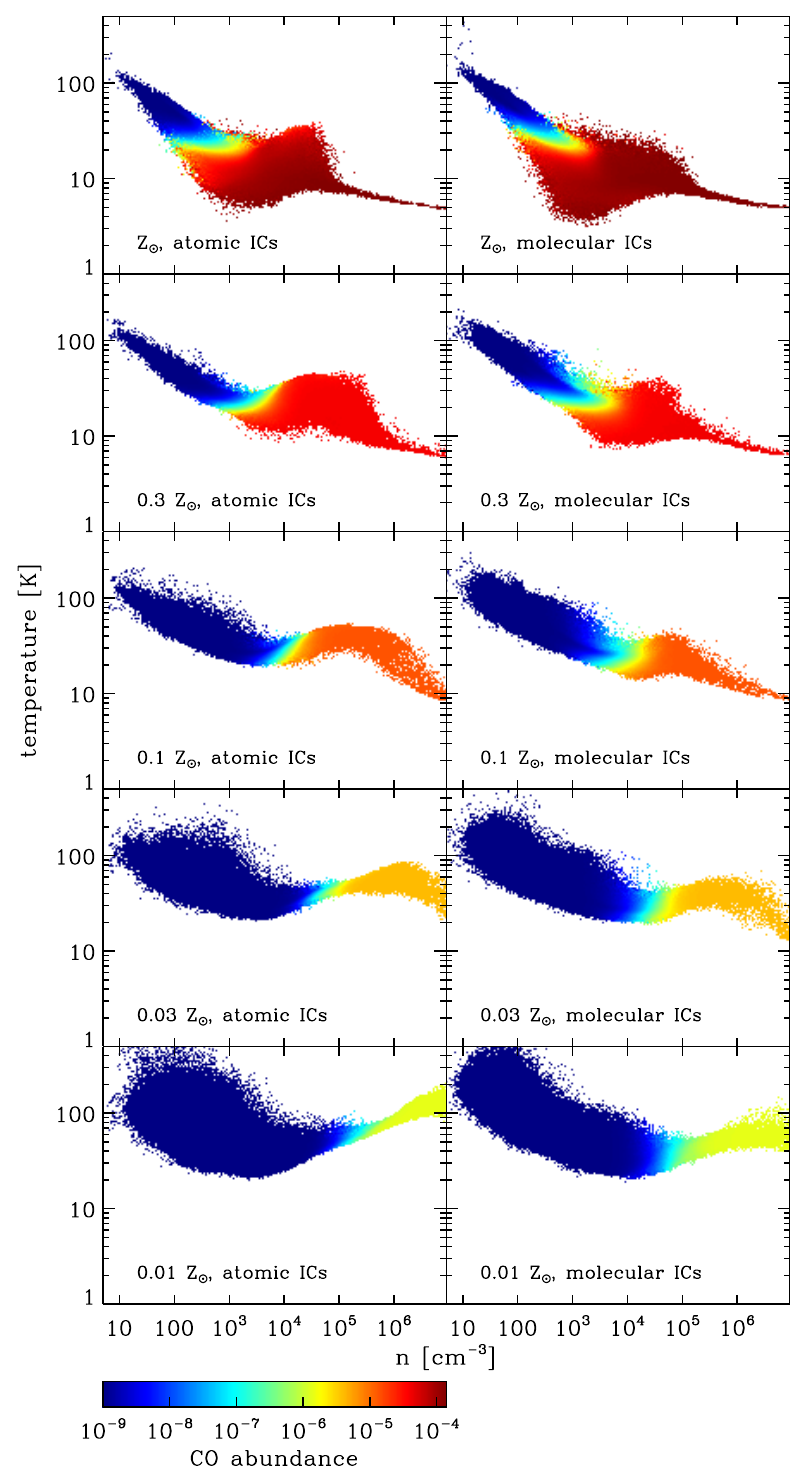}
\caption{Phase diagram showing the temperature of the gas as a function of its density,
colour-coded to indicate the CO abundance at each point in density-temperature space.
\label{fig:nt}}
\end{figure}

The phase diagrams plotted in Figure~\ref{fig:nt} also show us how the temperature distributions of our
simulated clouds vary with metallicity.  In our solar metallicity
clouds, we can see that there are three distinct regimes. At densities $n < 10^{3} \: {\rm cm^{-3}}$,
the cooling of the gas is dominated by C$^{+}$ fine structure emission, while the heating is dominated
by the photoelectric effect. The balance between these two processes yields a characteristic 
temperature that is a steeply decreasing function of the density, declining from $T \sim 100$~K at 
$n \sim 30 \: {\rm cm^{-3}}$ to $T \sim 20$~K at $n \sim 1000 \: {\rm cm^{-3}}$. At higher densities,
$10^{3} < n < 10^{5} \: {\rm cm^{-3}}$, there is considerable scatter in the temperature distribution.
This scatter is due in large part to the fact that the dominant coolant in this regime is CO, and the
effectiveness with which this can cool the gas depends to a large extent on the local optical depth 
of the CO $J = 1 \rightarrow 0$ rotational emission line \citep[see also the discussion of this point
in][]{gc12b}. In the run starting from initially atomic gas, an additional contribution to the scatter
in this density regime comes from the influence of H$_{2}$ formation heating, which is responsible
for the pronounced bulge in the distribution at $T \sim 20$--30~K and $n \sim 3 \times 10^{4} \:
{\rm cm^{-3}}$. Finally, at densities $n > 10^{5} \: {\rm cm^{-3}}$, the scatter in the temperature 
distribution abruptly vanishes, as the gas temperature becomes strongly coupled to the dust temperature.

If we now examine what happens as we reduce the metallicity, we see that the temperature
distribution tends to flatten. The temperature of the lowest density gas increases from roughly
100~K in the solar metallicity case to roughly 200--300~K in the ${\rm Z} = 0.01 \: {\rm Z_{\odot}}$
run, but the efficient cooling that the fine structure lines of C$^{+}$ and O provide at these 
temperatures even in the low-metallicity runs prevents the temperature from rising much further.
At the high density end of the distribution, however, the influence of metallicity is much more
pronounced. The timescale on which energy is transferred between the gas and the dust scales
with metallicity and density as $t_{\rm gd} \propto n^{-1} ({\rm Z}/{\rm Z_{\odot}})^{-1}$, 
and the gas and dust temperatures become strongly coupled only once this timescale becomes 
shorter than the local dynamical time, which in this case is the free-fall time of the gas. Therefore,
as Z decreases, the density at which the dust and gas temperatures become strongly coupled
increases. Consequently, in the lower metallicity clouds, much of the dense gas must rely on
CO cooling rather than dust cooling. However, CO cooling also becomes increasingly ineffective as
we move to lower metallicities, as the CO abundance declines. On the other hand, the rate at
which the dense gas is heated by the photoelectric effect {\em increases} as we decrease Z,  
owing to the consequent reduction in the mean extinction of the gas.\footnote{In optically thin
regions, reducing the metallicity {\em reduces} the photoelectric heating rate, since there is
less dust to absorb UV photons and emit photo-electrons. However, in high column density
regions, this effect is less important than the reduction in the mean extinction, as the latter 
leads to an exponential increase in the heating rate with decreasing Z, up to the point at which
the gas becomes optically thin.} The net effect is to increase
the temperature of the densest gas from roughly 5~K in the solar metallicity case to anywhere
between 50~K and 100~K in the ${\rm Z} = 0.01 \: {\rm Z_{\odot}}$ case. This increase in the 
temperature of the dense gas makes it much harder to create high density substructure within the cloud, 
whether through the action of turbulence or that of gravity, explaining why it takes longer for the cloud 
to start forming stars. 

The increase in the gas temperature will also lead to an increase in the Jeans mass. If we compare
the temperatures in the runs at a density of $10^{4} \: {\rm cm^{-3}}$, characteristic of most prestellar
cores, then we see that there is roughly a factor of two increase in the temperature between the
solar metallicity case and the $0.01 \: {\rm Z_{\odot}}$ case, corresponding to an increase in the
Jeans mass by about a factor of three. If we look at the densest gas in our simulations, we see an 
even stronger effect. The change in the Jeans mass that occurs as we reduce the metallicity could
potentially have a pronounced influence on the initial mass function of the stars that form,
although the limited mass resolution of our simulations does not allow us to address this issue with 
confidence.  Nevertheless, it is important to realise that the Jeans mass in the dense gas remains
much smaller than the mass of the cloud. Since the cloud itself also remains gravitationally bound
in even the lowest metallicity run, the gravitational collapse of part or all of the gas within it, and
the consequent formation of stars, is an inevitable outcome of the evolution of the cloud.

\subsection{CO emission}
\label{observe}
The results of the previous sections demonstrate that the CO content of the low
metallicity clouds is much smaller than the CO content of the solar metallicity clouds.
It is therefore reasonable to assume that the lower metallicity clouds will produce much
less emission in the $J = 1 \rightarrow 0$ rotational transition of $^{12}$CO, the most
commonly used observational tracer of molecular gas. To confirm this expectation,
we have computed emission maps for this transition for each of our model clouds using
the RADMC-3D radiative transfer code.

\begin{figure*}
\centerline{
\includegraphics[width=5.6in]{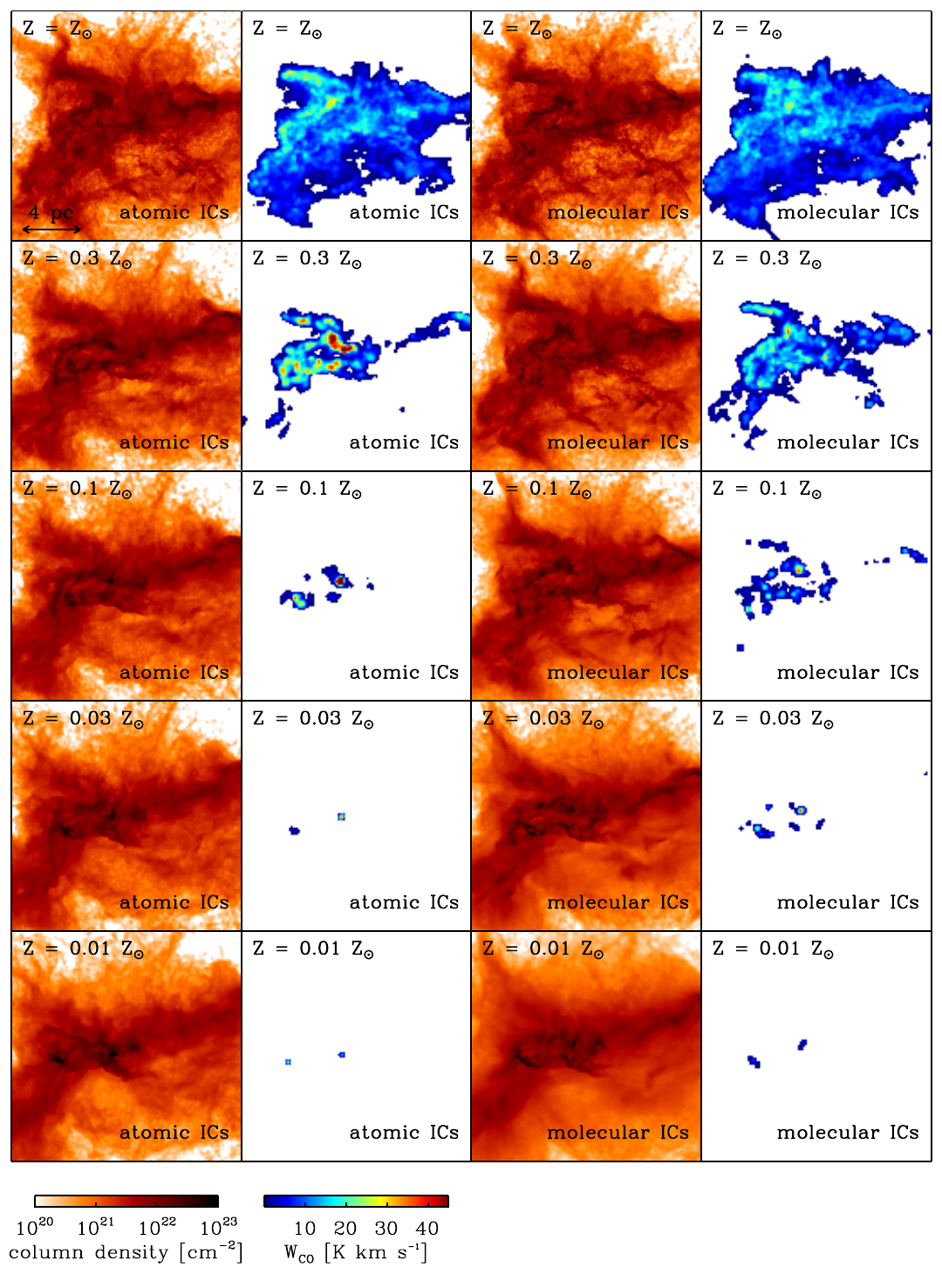}
}
\caption{Maps of column density (first and third columns) and integrated intensity in the 
$J = 1 \rightarrow 0$ rotational transition of $^{12}$CO (second and fourth columns) for
each of the simulations. The maps show a region of side length 16.2~pc that includes
roughly 80\% of the total cloud mass, but almost all of the CO emission. The CO integrated
intensity maps were produced using the RADMC-3D radiative transfer code, as described
in the text.  \label{fig:coemiss}}
\end{figure*}

In Figure~\ref{fig:coemiss}, we show velocity-integrated emission maps of the central region of 
each of our model clouds at a point just before the onset of star formation. In each case, we have
computed the emission arising from a cubic region of side length 16.2~pc centered on 
the cloud\footnote{Note that we expect very little CO 
emission to come from gas outside of this volume even in the solar metallicity case.}
which was discretised onto a $128^{3}$ grid. The CO level populations were computed
using the large velocity gradient (LVG) approximation, and we account for small-scale,
unresolved motions by including a uniform microturbulent contribution 
$v_{\rm mtb} = 0.2 \: {\rm km} \: {\rm s^{-1}}$ to the line broadening. Further details of the 
operation of the code can be found in \citet{shetty11a,shetty11b}.

The emission maps in Figure~\ref{fig:coemiss} demonstrate that as we reduce the metallicity
of the gas, the CO emission produced by the cloud does indeed decrease. The diffuse CO
emission that traces the majority of the cloud volume in the solar metallicity runs is quickly
lost as we move to lower metallicities, but the densest gas continues to produce distinct 
emission peaks down to very low metallicities. The peak value of the CO velocity-integrated 
intensity, $W_{\rm CO, max}$, does not show any clear dependence on the metallicity of the 
gas (Table~\ref{tab:co}), but the mean intensity,  $W_{\rm CO, mean}$, decreases rapidly as the
metallicity decreases. The mean intensity also tends to be smaller in the runs that start with 
atomic hydrogen rather than molecular hydrogen, although this is not a large effect in most of
the runs.

We have also computed the mean X-factor for each region, using the following definition:
\begin{equation}
X_{\rm CO} = \frac{N_{\rm H_{2}, mean}}{W_{\rm CO, mean}},  \label{xcoeq}
\end{equation}
where $N_{\rm H_{2}, mean}$ is the mean value of the H$_{2}$ column density.
We list the resulting values in Table~\ref{tab:co}, in units of the canonical value for 
Galactic GMCs, X$_{\rm CO, gal} = 2 \times 10^{20} \: {\rm cm^{-2}} \: ({\rm K \: km \: s}^{-1})^{-1}$
\citep[see e.g.][]{dame01}. We see that as we reduce the metallicity of the gas, 
X$_{\rm CO}$ increases, but that the strength of this effect depends on the initial
chemical state of the gas. In the runs that start with their hydrogen in molecular form,
the mean H$_{2}$ column density of the cloud remains large at low metallicities and
X$_{\rm CO}$ shows a strong sensitivity to metallicity, driven by the substantial decrease 
in $W_{\rm CO, mean}$ that occurs at low Z. On the other hand, the low metallicity runs 
that start with their hydrogen in atomic form have much lower H$_{2}$ column densities
than their higher metallicity counterparts, and hence show less variation in X$_{\rm CO}$,
as the reduction in $W_{\rm CO, mean}$ is offset by the reduction in the mean H$_{2}$
column density. The behaviour of real clouds probably lies somewhere between these
two extreme cases.

\begin{table}
\caption{CO intensities and X$_{\rm CO}$ \label{tab:co}}
\begin{tabular}{lccc}
\hline
Run & $W_{\rm CO, max}$ & $W_{\rm CO, mean}$ & X$_{\rm CO}$ / X$_{\rm CO, gal}$ \\
\hline
Z1-M & 25.7 & 3.46 & 2.06 \\   
Z1-A & 29.9 & 3.23 & 1.53  \\ 
\hline
Z03-M & 37.0 & 1.34 & 4.76 \\  
Z03-A & 53.5 & 1.27 & 1.97 \\  
\hline
Z01-M & 37.7 & 0.217 & 22.6 \\ 
Z01-A & 98.2 & 0.144 & 4.99 \\
\hline
Z003-M & 38.7 & 0.045 & 66.3 \\ 
Z003-A & 32.8 & 0.016 &  10.0 \\ 
\hline
Z001-M & 8.16 & 0.0068 & 306.7  \\ 
Z001-A & 25.6 & 0.0106 & 8.27 \\ 
\hline
\end{tabular}
\medskip
\\
{\bf Note:} $W_{\rm CO, max}$ is the maximum value of the CO velocity-integrated intensity, 
while $W_{\rm CO, mean}$ is the mean value, averaged over all of the lines of sight 
considered in the calculation. Both values have units of K km s$^{-1}$. X$_{\rm CO}$
is the mean conversion factor between CO intensity and H$_{2}$ mass, and 
X$_{\rm CO, gal} = 2 \times 10^{20} \: {\rm cm^{-2}} \: ({\rm K \: km s}^{-1})^{-1}$
is the canonical Galactic value of X$_{\rm CO}$ \citep{dame01}.
\end{table}

Another issue that should be borne in mind when considering how X$_{\rm CO}$ varies
with metallicity is that the values that we obtain for our low metallicity clouds are
highly time-dependent. The CO emission of these clouds is dominated by emission 
from a few dense, self-gravitating regions, and we would expect  $W_{\rm CO, mean}$
for these clouds to be much lower if we were to observe them at an earlier time in their
evolution, before the onset of gravitational collapse in these regions. 

Looking at the column density projections of the clouds (also
shown in Figure~\ref{fig:coemiss}), we see that as we decrease the metallicity, the
structure of the cloud changes. The gas becomes more centrally condensed as we
decrease Z, and loses much of the substructure that is present in the solar metallicity
run. This behaviour is a consequence of the higher gas temperatures found in the
lower metallicity runs. A higher temperature implies a higher sound-speed and hence
a lower Mach number for the turbulence, with the result that the turbulence is able to
generate less structure in the density field \citep[see e.g.][]{pnj97,pvs98,pfb11,mol12}. 
It also leads to a higher characteristic Jeans mass, making it less likely that the overdense 
regions generated by the turbulence will be gravitationally bound.

\subsection{Environmental sensitivity}
\label{environ}
{The simulations that we have discussed in detail above were performed using only a single, fixed
value for the strength of the interstellar radiation field and for the cosmic ray ionization rate. In reality,
both of these values will vary somewhat from region to region within a galaxy, and
from galaxy to galaxy. For example, we would expect the UV radiation field strength to be higher in
a low metallicity dwarf galaxy than in a high metallicity spiral, given the same surface density of 
star formation in both systems, owing to the smaller amount of dust extinction in the lower metallicity
system. 

Therefore, in order to establish the extent to which the conclusions that we draw in this paper depend 
on our assumptions regarding the strength of the ISRF and the cosmic ray ionization rate, we have
performed several simulations in which these values were varied. To prevent the number of simulations
from becoming completely impractical, we considered only the two extreme cases where ${\rm Z = Z_{\odot}}$
and ${\rm Z} = 0.01 \: {\rm Z_{\odot}}$, and in both cases adopted molecular initial conditions. We performed
simulations in which the strength of the ISRF was increased or decreased by a factor of ten (hereafter
referred to with labels of the form Z$n$-G10 and Z$n$-G01, respectively, where $n$ denotes the metallicity,
as before)\footnote{We remind
the reader than in most of our runs, the strength of the ultraviolet portion of the radiation field is
$G_{0} = 1$ in units of the \citet{dr78} field, while the longer wavelength portions of the field are taken from
\citet{bl94}. Our additional runs therefore correspond to runs with $G_{0} = 10$ and $G_{0} = 0.1$,
respectively, with a similar change also being made at longer wavelengths.} 
and simulations in which $\zeta_{\rm H}$ was increased or decreased by a factor of ten 
(hereafter referred to with labels of the form Z$n$-CR10 and Z$n$-CR01, respectively). In the runs in which 
$\zeta_{\rm H}$ was varied, the other cosmic ray ionization rates were also varied so as to keep their ratios 
with $\zeta_{\rm H}$ unaltered.

In our present study, we do not explore the effect of more extreme changes in the radiation field strength
or cosmic ray ionization rate, such as may be expected in starburst environments. We plan to address this
issue in future work.

\subsubsection{Effects of varying the strength of the ISRF}
\label{ISRF}
In the upper panel of Figure~\ref{fig:sfrate_env}, we show how increasing or decreasing the strength of
the interstellar radiation field affects the star formation rate in our model clouds. In the solar metallicity case,
we see that changes in the strength of the ISRF have very little effect -- all three of our model clouds begin
forming stars at almost the same time, give-or-take 0.1~Myr. The fact that we do not see
a strong dependence of the star formation rate on the strength of the ISRF in these runs is easy to understand,
as in these clouds, gravitationally-bound pre-stellar cores form only in regions with high dust extinctions.
The thermal state of the gas in the cores is therefore insensitive to changes in the strength of the ultraviolet
portion of the ISRF, as UV photons do not penetrate into these regions. It is sensitive to changes in the mid-IR and
far-IR portions of the ISRF, but in this case the dust temperature is only a very weak function of the radiation
field strength. For the dust model adopted in our simulations, $T_{\rm d} \propto U_{\rm IR}^{1/6}$, where
$U_{\rm IR}$ is the radiation energy density in the mid-IR and far-IR, and so changes of an order of magnitude in
$U_{\rm IR}$ lead to at most a 50\% change in the dust temperature, which is not enough to significantly 
affect the star formation process.\footnote{It may have some effect on the resulting stellar IMF, but an 
investigation of this point lies outside the scope of this paper.}

\begin{figure}
\includegraphics[width=3.2in]{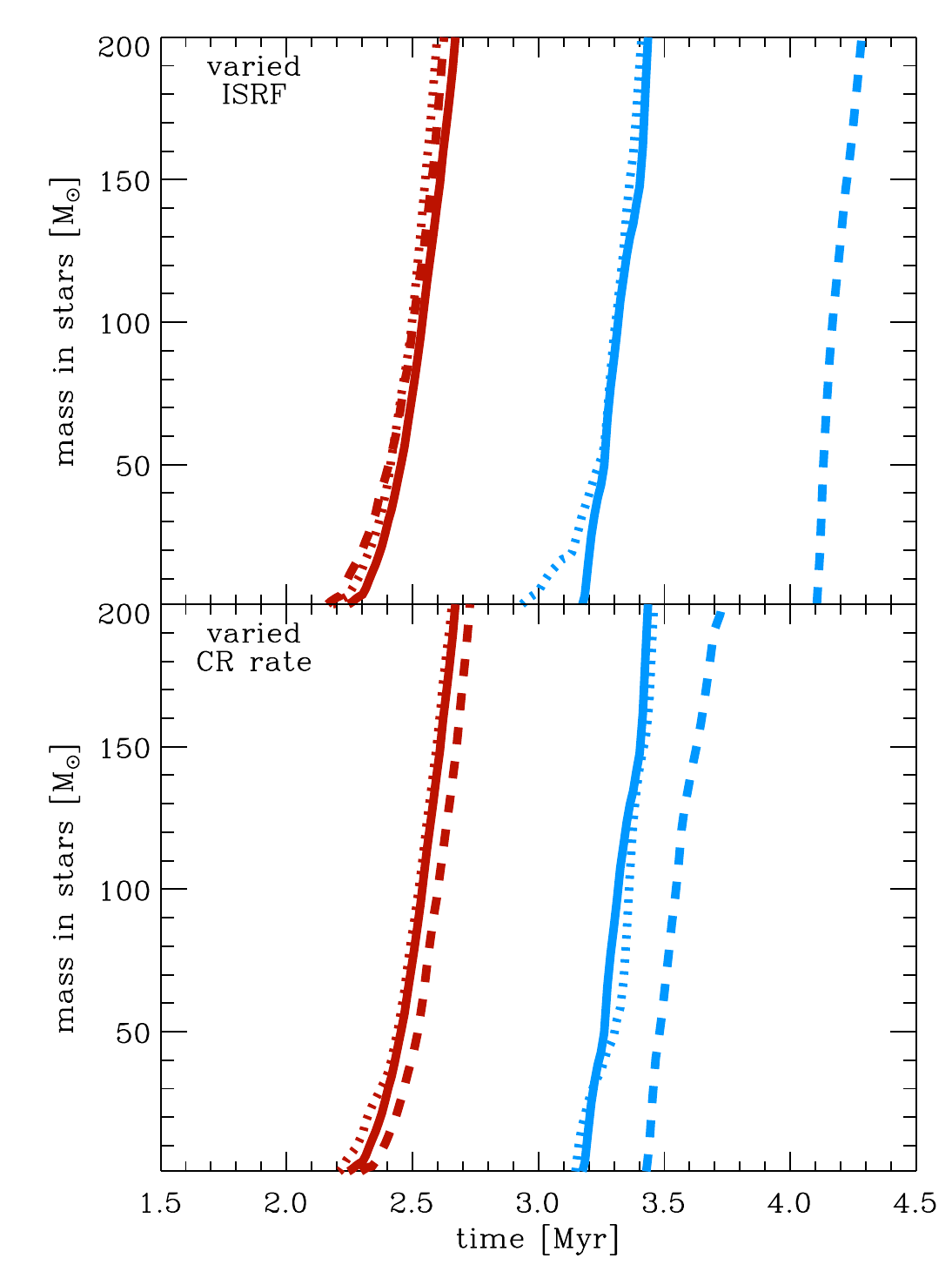}
\caption{{\it Upper panel}: Mass in sinks, plotted as a function of time, for runs Z1-M (left-hand solid line), 
Z1-G01 (left-hand dotted line), Z1-G10 (left-hand dashed line), Z001-M (right-hand solid line), 
Z001-G01 (right-hand dotted line) and Z001-G10 (right-hand dashed line). 
{\it Lower panel}: The same quantity, but for runs Z1-M (left-hand solid line), Z1-CR01 (left-hand dotted line), 
Z1-CR10 (left-hand dashed line), Z001-M (right-hand solid line), 
Z001-CR01 (right-hand dotted line) and Z001-CR10 (right-hand dashed line).
\label{fig:sfrate_env}}
\end{figure}

In our ${\rm Z} = 0.01 \: {\rm Z_{\odot}}$ runs, the effect of changing the strength of the ISRF is more 
pronounced. Reducing the radiation field strength by an order of magnitude has little effect on the
star formation rate, but increasing it by an order of magnitude delays the onset of star formation by
almost $1 \: {\rm Myr}$. However, once star formation begins in run Z001-G10, it proceeds at 
about the same rate as in the other runs.

In Table~\ref{chemtab2}, we show how the chemical state of the gas at the onset of star formation varies as we
vary the strength of the ISRF. In the solar metallicity runs, we see that although the fraction of the
hydrogen that is in the form of H$_{2}$ decreases as we increase the radiation field strength, all
three of our simulated clouds remain primarily molecular. The effect on the carbon chemistry is more
pronounced, with the mass fraction of carbon in the form of CO, $F_{\rm CO}$, decreasing by around
a factor of two for each factor of ten increase in the radiation field strength. The amount of C$^{+}$ in
the clouds also changes substantially, while the amount of atomic carbon changes by less than 50\%.

Once again, however, changes in the radiation field strength have a much larger effect in the low
metallicity runs than in the solar metallicity runs. Decreasing
the strength of the field by a factor of ten dramatically reduces the amount of H$_{2}$ that is destroyed 
by photodissociation, meaning that the hydrogen in the cloud remains largely in molecular form at the
onset of star formation. On the other hand, increasing the radiation field strength leads to an almost 
complete loss of H$_{2}$ from the gas. The effect on the CO is less pronounced, as the CO is concentrated
in dense gas that is less susceptible to the effects of photodissociation, but even in this case we see a
factor of two change in the CO mass fraction for each factor of ten change in the radiation field strength.

\begin{table}
\caption{Chemical state of the gas in runs with different ISRF strengths and cosmic
ray ionization rates \label{chemtab2}}
\begin{tabular}{lcccc}
\hline
Run & $F_{\rm H_{2}}$ & $F_{\rm C^{+}}$ & $F_{\rm C}$ & $F_{\rm CO}$ \\
\hline
Z1-M & 0.883 & 0.601 & 0.084 & 0.315 \\ 
\hline
Z1-G10 & 0.570 & 0.804 & 0.064 & 0.132 \\  
Z1-G01 & 0.989 & 0.291 & 0.105 & 0.604 \\ 
\hline
Z1-CR10 & 0.871 & 0.642 & 0.166 & 0.192 \\
Z1-CR01 & 0.886 & 0.593 & 0.076 & 0.331 \\ 
\hline
\hline
Z001-M & 0.260 & 0.990 & 0.003 & 0.007 \\ 
\hline
Z001-G10 & 0.004 & 0.995 & 0.001 & 0.004 \\  
Z001-G01 & 0.904 & 0.965 & 0.021 & 0.014 \\  
\hline
Z001-CR10 & 0.201 & 0.990 & 0.004 & 0.006 \\ 
Z001-CR01 & 0.269 & 0.992 & 0.002 & 0.006 \\ 
\hline
\end{tabular}
\medskip
\\
{\bf Note:} $F_{\rm H_{2}}$, $F_{\rm C^{+}}$, etc.\ are the same as in Table~\ref{chemtab}.
Values from runs Z1-M and Z001-M, performed using our standard values for the ISRF
and cosmic ray ionization rate are listed here for reference.
\end{table}

\subsubsection{Effects of varying the cosmic ray ionization rate}
\label{cosmic}
In the lower panel of Figure~\ref{fig:sfrate_env}, we show how the star formation rate varies as we increase or decrease
the cosmic ray ionization rate by an order of magnitude. In the solar metallicity case, we see that variations of an order
of magnitude in $\zeta_{\rm H}$ have almost no effect on the star formation rate. At low gas densities, heating due to
cosmic ray ionization is less important than photoelectric heating even in the simulation with the increased value
of $\zeta_{\rm H}$, and so variations in $\zeta_{\rm H}$ therefore have no significant effect on the temperature of the 
low density gas. At higher densities, photoelectric heating becomes ineffective owing to the increasing extinction, and
in this regime, cosmic ray heating can become important. However, CO line cooling and dust cooling are both very 
efficient in this dense gas, and so even an order of magnitude increase in the heating rate leads to only a very
small change in the temperature. The effect is therefore too small to have a significant impact on the ability of the
cloud to form stars.

In our low metallicity simulations, decreasing $\zeta_{\rm H}$ again has almost no effect on the star formation rate.
On the other hand, if we increase $\zeta_{\rm H}$ by a factor of ten, there is a clear effect -- the onset of star formation
is delayed, but only by around 0.2~Myr, corresponding to around 10\% of the cloud free-fall time. The cosmic rays have
a greater effect in this case than in the solar metallicity case because the low metallicity gas cools much less efficiently
than the solar metallicity gas. Nevertheless, the main change in the temperature structure occurs in dense gas that is
already self-gravitating or close to collapse, and the temperature change therefore has little influence on the star 
formation rate.

Looking at the chemical state of the clouds at the onset of star formation, we see that the effect of varying 
$\zeta_{\rm H}$ is much smaller than that of varying the strength of the ISRF. In our solar metallicity simulations,
decreasing $\zeta_{\rm H}$ by a factor of ten leads to minor increases in the H$_{2}$ fraction and CO fraction
(and corresponding decreases in the amount of C and C$^{+}$), but only at the level of a few percent. Increasing
$\zeta_{\rm H}$ by an order of magnitude has a stronger effect on the CO abundance, reducing it by around 40\%.
This is because in the run with higher $\zeta_{\rm H}$, more He$^{+}$ ions are produced, which then destroy CO
via the dissociative charge transfer reaction
\begin{equation}
{\rm CO} + {\rm He^{+}} \rightarrow {\rm C^{+}} + {\rm O} + {\rm He}.
\end{equation}
On the other hand, the amount of H$_{2}$ present in the cloud is barely affected, changing by less than 2\%.

In our low metallicity runs, changing  $\zeta_{\rm H}$ has a stronger effect on the H$_{2}$ content of the cloud,
decreasing it by almost a quarter in run Z001-CR10 compared to the value in run Z001-M. Since the formation
rate of H$_{2}$ in these clouds is considerably lower than in the solar metallicity case, it is unsurprising that
the cosmic rays have a larger effect. Looking at the abundances of the carbon species, however, we see that
they barely change. This is because the majority of the C and CO in these simulations is found in high
density, gravitationally collapsing gas, as we have already discussed in Section~\ref{chem_comp}, and cosmic 
ray ionization has little influence on the chemical content of these high density regions.
}

\section{Discussion}
Our simulations demonstrate that the metallicity of the gas has a surprisingly small effect
on the star formation rate on the scale of individual, gravitationally-bound clouds. Decreasing
the metallicity by two orders of magnitude, from ${\rm Z_{\odot}}$ to $0.01 \: {\rm Z_{\odot}}$,
delays star formation on this scale by at most a cloud free-fall time, but does not strongly affect
the rate at which stars form once star formation begins within the cloud. As we have seen in
Sections~\ref{chem} and \ref{chem_comp}, reducing the metallicity tends to make the dense 
gas within the cloud warmer, raising its Jeans mass and reducing the amount of dense substructure 
that can be created by the combined effects of turbulence and gravity. However, the Jeans mass 
remains considerably smaller than the cloud mass, and so gravitational collapse and the
formation of stars remains an inevitable outcome.

Our models also show that changing the metallicity has a strong effect on the CO content
of the clouds, in agreement with previous theoretical and numerical work 
\citep[see e.g.][]{mb88,sakamoto96,bell06,gm11, shetty11a}. As we decrease Z, the
CO becomes increasingly concentrated within the densest gas and the mean CO 
abundance sharply decreases. As a consequence, the mean intensity of the CO emission 
produced by the clouds also rapidly decreases, although the peak intensities remain
roughly similar in all of the models.  

If we take our computed values of $W_{\rm CO, mean}$ as a suitable proxy for the total
CO luminosity of the clouds, then we can show that the star formation rate per unit CO
luminosity increases by a factor of up to 400 as we decrease the metallicity from 
${\rm Z_{\odot}}$ to $0.01 \: {\rm Z_{\odot}}$ (see Table~\ref{tab:sfr}). Our results therefore
provide strong support for the hypothesis that the anomalously high values of the star
formation rate per unit CO luminosity found in many low metallicity systems are due 
primarily to a deficit in the amount of CO in these systems, rather than any fundamental
change in the star formation process. That said, we do find that at very low metallicities,
the star formation rate per unit H$_{2}$ mass increases (see Table~\ref{tab:sfr}), although
the strength of this effect depends strongly on the initial chemical composition assumed
for the clouds, {and on the strength that we adopt for the interstellar radiation field}.

\begin{table}
\caption{Normalised star formation rates \label{tab:sfr}}
\begin{tabular}{lcc}
\hline
Run & SFR$_{\rm CO}$ & SFR$_{\rm H_{2}}$ \\
\hline
Z1-M & 1.00 & 1.00 \\ 
Z1-A & 0.86 & 1.18 \\ 
\hline
Z03-M & 2.05 & 0.90 \\ 
Z03-A & 1.87 & 1.99 \\ 
\hline
Z01-M & 13.8 & 1.28 \\ 
Z01-A & 15.6 & 6.25 \\
\hline
Z003-M & 50.7 & 1.57 \\ 
Z003-A & 134 & 28.7 \\ 
\hline
Z001-M & 395 & 2.63 \\  
Z001-A & 191 & 66.8 \\   
\hline
\end{tabular}
\medskip
\\
{\bf Note:} SFR$_{\rm CO}$ is the star formation rate of the cloud per unit CO luminosity,
normalized to the value in run Z1-M. SFR$_{\rm H_{2}}$ is the star formation rate of the cloud 
per unit H$_{2}$ mass, normalized in a similar fashion.
\end{table}

How do we reconcile these results with recent work showing that galaxy formation models in
which the star formation efficiency declines with decreasing metallicity  do a far better job of 
reproducing observations of the Kennicutt-Schmidt relation and the star formation rate per unit 
mass in low-mass galaxies than models in which the efficiency is independent of metallicity?
(See e.g.\ the models of \citealt{gk10}, \citealt{kuhlen11} or \citealt{kd11}).
It is important to note at this point 
that there are at least two ways in which the metallicity of the gas could have a strong influence on 
the star formation rate that are not addressed by our current models. 

First, our models assume that the gas has already been assembled into a gravitationally-bound
cloud, and do not address how this actually occurred. However, it is not unreasonable to expect
that the process of cloud assembly may have some sensitivity to metallicity. For example, 
consider the classical picture of the two-phase neutral interstellar medium, as described in
detail in e.g.\ \citet{wolf95,wolf03}. Models of the two-phase medium show that if we reduce
both the dust and gas-phase metallicities, but keep all of the other environmental parameters
(e.g.\ the cosmic ray ionization rate or the strength of the interstellar radiation field) constant, 
then the minimum ISM density required in order to allow for the existence of a cold neutral
phase increases \citep[see e.g.\ Figure 6 in][]{wolf95}. Since the existence of a cold neutral
phase is an obvious pre-requisite for star formation, this result highlights one way in which
the metallicity of the gas may affect the star formation rate on large scales: by making it more
difficult to produce cold gas, we also make it more difficult to produce stars, and hence reduce
the star formation rate. This idea has some support from numerical models of turbulence in
the low metallicity ISM \citep[e.g.][]{walch11}, but deserves to be studied in more detail.

The second main way in which the metallicity of the gas may affect the star formation rate
is through the influence of stellar feedback. It is plausible that stellar feedback, particularly
in the form of UV radiation, may become more effective at suppressing star formation as
we reduce the metallicity of the gas, owing to the decrease in the cooling rate of the clouds
and their reduced ability to shield themselves from the UV radiation. {The results presented
in Section~\ref{ISRF} for an interstellar radiation field that was ten times stronger than our default
value give some support to this idea, as in this case we do find a delay in the onset of star
formation in our ${\rm Z} = 0.01 \: {\rm Z_{\odot}}$ simulation. However, the effect is relatively
small, suggesting that if we want to significantly reduce the star formation efficiency of the
cloud, a large enhancement in the radiation field strength will be necessary.}

On the other hand,
\citet{dib11} argue that stellar feedback should actually be {\em less} effective at lower
metallicities. In their model, they assume that stellar winds from massive stars are the most
important form of feedback, and these are well known to grow weaker with decreasing
metallicity \citep[see e.g.][]{kud02}, implying that stellar wind feedback will also become less effective.
Feedback models in which radiation pressure from massive stars plays a leading role
also predict that feedback will become less effective as Z decreases, owing to the 
reduction in the ability of the dust to trap the radiation (A.\ Kravtsov, private communication).
It is therefore safe to say that the jury is still out regarding the overall influence of  metallicity
on the effectiveness of stellar feedback, and whether the reduced efficacy of star formation
at low metallicities that is apparently required by galaxy formation models can be explained 
by a change in the effectiveness of stellar feedback.

\section{Conclusions}
We have shown in this paper that the star formation rate of individual, gravitationally bound
clouds does not have a strong dependence on the metallicity of the gas making up those 
clouds. Although changes in the metallicity lead to pronounced changes in the chemical
composition of the clouds and their temperature distribution, the effect on the star formation
rate is comparatively small: two orders of magnitude decrease in the metallicity delays the
onset of star formation by less than a single cloud free-fall time. 

Our results provide strong support for a picture in which the high values observed for the
star formation rate per unit CO luminosity in many low metallicity systems are caused primarily
by a large deficit in the amount of CO present in comparison to more metal-rich systems,
rather than by an increase in the star formation efficiency in the lower metallicity systems.
We also find evidence that the star formation rate per unit H$_{2}$ mass should increase with 
decreasing metallicity, although the strength of this effect depends quite strongly on the 
assumptions we make regarding the initial chemical composition of our model clouds,
{and on the strength of the incident UV radiation field}. Observational confirmation
of this effect will be difficult, because at metallicities ${\rm Z} > 0.1 \: {\rm Z_{\odot}}$ the size of 
the effect appears to be smaller than the typical observational uncertainty in the amount of 
H$_{2}$ present in the gas \citep[see e.g.][who conclude that their inferred H$_{2}$ column
densities are uncertain to within a factor of 2--3]{bol11}.
Nevertheless, efforts to test the model by determining
the star formation rate per unit H$_{2}$ mass in very low metallicity systems would be very
valuable, as the increase that our models predict distinguishes them from many other models 
of star formation in GMCs in which the star formation rate per unit H$_{2}$ mass remains
constant \citep[see e.g.][]{klm11}.

Another interesting result of our simulations is the demonstration that in low metallicity 
clouds, the CO abundance is highly time-dependent. CO forms efficiently in these clouds
only at very high gas densities, and these densities are reached only within gravitationally
collapsing pre-stellar cores. Prior to the onset of gravitational
collapse, the CO content of these clouds is typically very small, meaning that they will have
low average CO luminosities and high CO-to-H$_{2}$ conversion factors. Once gravitational
collapse begins, however, the CO content of the clouds increases significantly, leading to
higher CO luminosities and hence lower CO-to-H$_{2}$ conversion factors. One way of testing
this prediction would be to compare the values of X$_{\rm CO}$ derived for starless and
star-forming clouds in low metallicities galaxies: if our models are correct, we would expect 
the starless clouds to have systematically higher values of X$_{\rm CO}$ than the star-forming
clouds.

Finally, it is important to remember that our results do not by themselves imply that the
star formation rate in galaxies is independent of metallicity. If decreasing the metallicity of 
the ISM makes it harder to form gravitationally bound clouds, then it will tend to decrease
the galactic star formation rate even if -- as demonstrated here -- the rate at which stars form 
within the individual clouds is barely affected. Similarly, if stellar feedback becomes more effective at lower metallicity,
then one would again expect the galactic star formation rate to be smaller. A definitive
answer to the question of the influence of metallicity on the star formation rate within 
galaxies is unlikely to be obtained before we can model the whole cloud lifecycle, from 
assembly to dispersion. The results presented here represent a small but important step
towards this ultimate goal.

\section*{Acknowledgments}
The authors would like to thank C.~Dobbs, N.~Gnedin, R.~Klessen, A.~Kravtsov, M.~Krumholz, A.~Kritsuk, S.~Madden, 
M.~Norman, R.~Shetty and A.~Wolfe for stimulating discussions regarding aspects of the 
work presented in this paper. They would also like to thank the anonymous referee for
suggestions that helped to improve the paper. The authors acknowledge financial support from the 
Baden-W\"urttemberg-Stiftung for contract research via grant P-LS-SPII/18, and the 
Deutsche Forschungsgemeinschaft (DFG) via SFB project 881, ``The Milky Way System''
(sub-projects B1 and B2) and priority program 1573, ``Physics of the Interstellar Medium''.
The GADGET simulations discussed in this paper were performed on the {\em Ranger} 
cluster at the Texas Advanced Computing Center, using time allocated as part of 
Teragrid project TG-MCA99S024. The radiative transfer modelling discussed in 
Section~\ref{observe} was performed on the {\em Kolob} cluster at the University of Heidelberg, 
which is funded in part by the DFG via Emmy-Noether grant BA 3706.

\end{document}